\documentclass[9pt,twocolumn,twoside]{osajnl}
\journal{ol} 
\setboolean{shortarticle}{true} 
\title{Single-shot non-interferometric measurement of the phase transmission matrix in multicore fibers}
\usepackage{amsfonts,amsmath,bm,bbm}

\newcommand{\be}[0]{\begin{equation}}
\newcommand{\ee}[0]{\end{equation}}
\newcommand{\bea}[0]{\setlength\arraycolsep{2pt}\begin{eqnarray}}
\newcommand{\eea}[0]{\end{eqnarray}}
\newcommand{\ui}{\mathrm{i}}

\newcommand{\bu}{\mathbf{u}}

\newcommand{\bx}{{\bf x}}
\newcommand{\bX}{{\bf X}}

\author[1]{Siddharth Sivankutty}
\author[1]{Viktor Tsvirkun}
\author[2]{G\'{e}raud Bouwmans}
\author[2]{Esben Ravn Andresen}
\author[3]{Dan Oron}
\author[1]{Herv\'{e} Rigneault}
\author[1,4*] {Miguel A. Alonso}
\affil[1]{Aix Marseille Univ., CNRS, Centrale Marseille, Institut Fresnel, UMR 7249, 13013 Marseille, France}
\affil[2]{PhLAM CNRS, IRCICA, Universit\'{e} Lille 1, 59658 Villeneuve d'Ascq Cedex, France}
\affil[3]{Department of Physics of Complex Systems, Weizmann Institute of Science, Rehovot, Israel}
\affil[4]{The Institute of Optics, University of Rochester, Rochester NY 14627}
\affil[*]{Corresponding author: miguel.alonso@fresnel.fr}
\dates{Compiled \today}
\ociscodes{060.2430, 060.2350, 170.2150, 180.5810, 070.6120, 110.1080}
\doi{\url{http://dx.doi.org/10.1364/ol.XX.XXXXXX}}

\begin{abstract}

A simple technique for far-field single-shot non-interferometric determination of the phase transmission matrix of a multicore fiber with over 100 cores is presented. This phase retrieval technique relies on the aperiodic arrangement of the cores.
\end{abstract}

\setboolean{displaycopyright}{true}
\begin{document}
\maketitle
While conventional endoscopes have demonstrated excellent imaging capabilities in terms of resolution, they suffer in terms of invasiveness due to the relatively bulky components placed on their distal tip. This fact restricts their application to hollow organs, ruling out imaging sensitive organs or conducting long-term studies where the probe needs to remain \textit{in-situ}. Amongst other strategies to overcome this size restriction \cite{gmitro1993confocal,gora2013tethered,tearney1996scanning}, fiber-based lensless endoscopes  have emerged as promising candidates that combine high-quality optical imaging with minimal invasiveness. One of the main problems with this type of device is that image transmission through a fiber is affected by modal mixing and dephasing \cite{vcivzmar2012exploiting}, but this problem can be addressed (either optically or computationally) if the complex transmission matrix (TM) of the fiber modes is known. 

This approach has been successfully demonstrated in fiber-based lensless endoscopes, with imaging modalities ranging from scanning microscopy, nonlinear microscopy, wide-field imaging, and even the generation of light sheets through a fiber \cite{vcivzmar2012exploiting,andresen2013toward,andresen2013two,Sivankutty2016a,psaltis_multimode,choi_multimode,ploschner2015multimode}.

For the transition of these lensless endoscopes from the optical table to a point-of-care destination, minimal complexity and a very high degree of robustness are required. In particular, multi-core fibers (MCF, also referred to as imaging fiber bundles), offer a great reduction in the complexity of the instrumentation \cite{andresen2013toward,thompson2011adaptive}. MCFs have been used in a configuration where each core of the fiber acts as an imaging pixel, performing an intensity mapping between the two ends of the fiber. Recent demonstrations \cite{Sivankutty:16} indicate that sampling or controlling the wavefront offers more prospects in terms of artifact-free imaging and 3D resolution. MCFs offer several degrees of freedom at the manufacturing process, where control can be exercised in parameters such as mode density and coupling strength, resulting in an direct tailoring of the TM properties. For operational advantages, these are engineered such that only the diagonal elements of the TM are significant, leading to an infinite memory effect \cite{Sivankutty:16}, reduced modal dispersion and reduced bending sensitivity \cite{Tsvirkun:17}. 
Another structural degree of freedom that can be exploited is the spatial distribution of the cores. Recently \cite{Sivankutty:18a}, a MCF was produced whose cores are arranged as a golden spiral \cite{gabrielli2016aperiodic}, an aperiodic configuration with roughly uniform density. Amongst other advantages, this arrangement leads to an important reduction of side lobes when focusing light at the distal end.
A key challenge in the deployment of lensless endoscopes is to achieve focusing at the fiber's distal end through non-interferometric methods such as iterative optimization \cite{caravaca2013real} or phase retrieval to obtain the fiber's TM \cite{Kogan:17}. However, the characterization of the TM typically requires a sequential interferometric process \cite{andresen2013toward}. In this letter, we propose a computationally inexpensive phase retrieval technique based on a single measurement of the speckle pattern emanating from a non-periodic MCF that is robust to minor amplitude changes.

Consider a MCF whose $N$  cores are uncoupled and have locations at the end facet given by the known transverse coordinates $\bx_n$ for $n=1,...,N$. We assume that the polarization emerging from all cores is the same, so we treat the field as scalar. The far-field intensity pattern generated by this fiber in terms of the transverse direction cosines $\bu$ is given by
\begin{equation}
I(\bu) = A(\bu)\left|\sum_{n=1}^{N}E_n~\exp[\ui(k\bu\cdot\bx_n+ \phi_n)]\right|^2,
\label{eqn:FF}
\end{equation}
where $k$ is the light's wavenumber, $A(\bu)$ is the far-field radiation pattern of the cores, and $E_n$ and $\phi_n$ are the (real) amplitude and phase for core $n$, respectively.  
A well localized intensity peak around a given direction $\bu$ requires that all contributions in Eq.~(\ref{eqn:FF}) are in phase \cite{Sivankutty:16}. Since the core positions and the mode structure of the light exiting a single core are known, the measurement of the TM reduces to the determination of $E_n$ and $\phi_n$. The amplitudes $E_n$ are easy to measure non-interferometrically, and they tend to be more uniform, less sensitive to manipulation of the MCF, and less critical to the achievement of a sharp focus than $\phi_n$. We then focus on retrieving the phases $\phi_n$, measured with respect to the phase of the central core, $\phi_1=0$. This problem has been examined through alternating projection (AP) algorithms for an aperiodic MCF \cite{Kogan:17}. 

Phase retrieval problems in 2D with source and far-field intensity measurements tend to lead to unique solutions \cite{fienup_review}, at least when important assumptions such as sparsity or low noise can be made.  A key point \cite{Kogan:17,Sivankutty:16} is the nature of the autocorrelation of the field distribution at the source plane in a MCF.  For a perfectly periodic MCF, this autocorrelation is sparse and concentrated around discrete points, since different peaks due to the correlations of several pairs of points (with equal separation vector) overlap. AP algorithms can be susceptible to the sensor's limited dynamic range and readout noise, to minor aberrations in the optical system, and to small changes in the intensities of the emitting cores, especially when a large number of cores ($\geq 10$ ) is measured simultaneously \cite{Kogan:17}. Note that amplitude modulations are inevitable in experimental scenarios, so phase retrieval techniques with limited sensitivity to changes in relative amplitudes are mandatory. These developments would ultimately aid in the implementation of an interferometer-free characterization and real-time monitoring of phase distortions in MCF-based lensless endoscopy. Phase retrieval in combination with compressive sensing has been already used to measure the TM of highly scattering media, albeit with multiple measurements \cite{dremeau2015reference}. As is shown in what follows, many of these problems are alleviated when the MCF is aperiodic, since the overlap of the correlation points can be reduced significantly, making them a better candidate for phase retrieval approaches in lensless endoscopy \cite{Kogan:17}. A similar strategy has been used in astronomy \cite{AMI}.

\begin{figure}[t]
  \centering
  \includegraphics[scale=1]{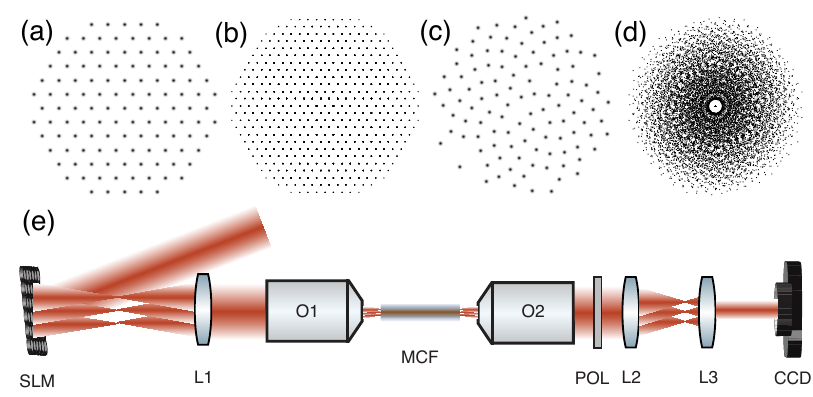}
  \caption{(a) Arrangement of the cores in a golden spiral. (b) Corresponding location of the auto-correlation peaks, scaled to $1/2$ of part (a). (c) Diagram of the optical setup.
\label{fig1}}
\end{figure}

The aperiodic MCF studied here is composed of 120 cores centered at the points $\bx_n=(\rho_n\cos\theta_n,\rho_n\sin\theta_n)$ arranged as a golden spiral [see Fig.~\ref{fig1}(a)] according to
\begin{align}
  \rho_n &= \Lambda \sqrt{n},\,\,\,\,\,\,\,\,
  \theta_n =   n\pi (3-\sqrt{5}),
\label{Eqn:Fermat}
\end{align}
where $\Lambda \approx 11.8 ~\mu$m is a parameter that regulates the typical inter-core spacing. The individual mode field diameter (MFD) of each core is $3.2 ~\mu$m. The fabrication and imaging properties of this golden-spiral MCF are  detailed in \cite{Sivankutty:18a}.
A simplified view of the setup used to project random phases into each of these cores and to record the far-field intensity pattern is depicted in Fig.~\ref{fig1}(c). Light from a CW laser (1053 $\pm$1 nm, IPG Lasers) is 
made to pass through
a liquid crystal SLM (X10468, Hamamatsu) which displays a phase profile corresponding to a lenslet array, generating an array of focal spots with controllable phases. An optical system (not shown) matches the size and divergence of these focal spots to the MCF cores to maximize the coupling efficiency. On the distal end of the MCF, an objective lens \textbf{O2}, (Olympus 20x, 0.48 NA) is used to access the far field of the MCF facet, and a relay system with 1.5x magnification (\textbf{L2}-\textbf{L3}) images this far field on a 8-bit CMOS detector (DCC1545, Thorlabs). Since the polarization state of light emanating from each fiber core is scrambled \cite{Sivankutty:16b}, a polarizer is used to ensure interference and enhance the contrast of the generated speckle. 

In order to obtain reference values for validating the results of the proposed method, we performed an independent interferometric calibration of the MCF's phase distortions, based on the far-field intensity measurement of each individual core with respect to the central one. Each of these fringe patterns was Fourier-transformed and the phase at the associated spatial frequency was obtained as in \cite{andresen2013two}, leading to reference phase measurements with an accuracy of $\pi/10$. This level of error stems predominantly from the laser and environmental fluctuations. A pre-selection of the cores useful for our measurements was also made during this process, causing us not to use 8 of the 120 cores for reasons of poor SNR arising from the polarization filtering. These calibration measurements were also used to measure the actual position of the cores, shown in Fig.~\ref{fig1}(a). 

\begin{figure}[h] 
\centering 
\includegraphics[scale = .4]{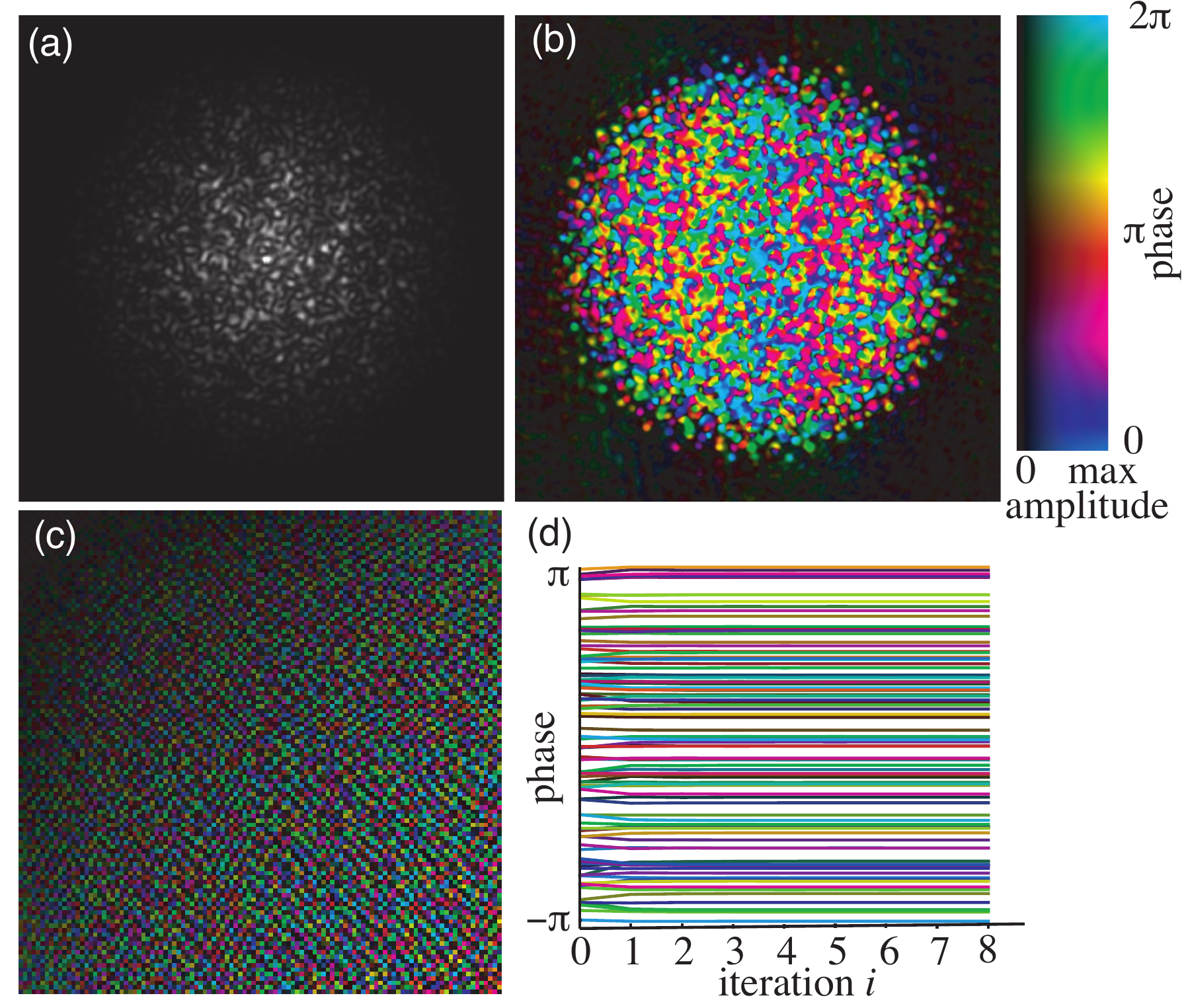} 
\caption{(a) Measured far-field speckle pattern for one of the 15 random phase realizations. (b) The corresponding complex autocorrelation function (partially saturated to show the main features). (c) Elements of the matrix $\mathbb{P}$. (d) Recovered values of the 112 phases for eight iterations.} 
\label{fig2} 
\end{figure}

The actual single-shot phase measurements were implemented by projecting random but known phase offsets into each of the 112 cores by using the SLM. Since the phase values were drawn randomly, speckle patterns in the far-field were obtained when all active 112 cores were used. We conducted 15 independent trials with different projected phases and their corresponding far-field speckle patterns were recorded, each of which served as the starting point for the phase retrieval method described in what follows. One of these speckle patterns is shown in Fig.~\ref{fig2}(a).

The key for the simple phase retrieval scheme used here is the known, aperiodic core arrangement, together with the fact that the mode field diameter is relatively small compared with the core spacing (by a factor of 0.27). This means that the spatial field autocorrelation at the plane of the fiber output is composed of many localized spots, each providing information about the correlation of the fields emerging from a given pair of cores. The aperiodic core distribution and small mode diameter help reduce the overlap of these correlation spots, as is discussed in what follows. 

The autocorrelation is found through the Fourier transform of the far-field intensity measurement. Since the far-field speckle pattern is asymmetric, the autocorrelation is a complex function, as represented in Fig.~\ref{fig2}(b) for the speckle pattern in Fig.~\ref{fig2}(a), where phase is encoded as color. The significant values of the autocorrelation are concentrated around the specific discrete locations given by the differences between any two core positions, $\bX_{n,m}=\bx_n-\bx_m$, shown in Fig.~\ref{fig1}(b). Given the aperiodicity of the core locations, the points $\bX_{n,m}$ do not coincide, but the density of these points does tend to increase as one approaches the origin, and this implies that the corresponding correlation spots overlap more significantly in that region. 

In order to retrieve the unknown phases $\phi_n$, a matrix is created whose elements are the phases $\Phi_{n,m}$ of the autocorrelation at the known points $\bX_{n,m}$ central to each spot. It is convenient for the measured intensity pattern to be approximately centered so that the phase over the extension of each spot in the autocorrelation is fairly uniform.
Notice that $\Phi_{n,m}=-\Phi_{m,n}$, given the fact that the autocorrelation is the Fourier transform of a real distribution. 
Ideally, $\Phi_{n,m}$ should coincide with $\phi_n-\phi_m$, where $\phi_n$ are the unknown phases of the cores. Therefore, these unknown phases can be estimated by minimizing the merit function
\be
\mu=\sum_{m,n=1}^N W_{n,m}^2\left|\exp(\ui\Phi_{n,m})-\exp[\ui(\phi_n-\phi_m)]\right|^2,
\ee
where the factors $0\le W_{n,m}\le1$ are used to penalize correlation points $\bX_{n,m}$ where the phase $\Phi_{n,m}$ is suspected not to correspond accurately to the phase difference between the field at cores $n$ and $m$ due to overlap with the spots for the correlation of other core pairs. Different criteria can be used for choosing these weights, such as the actual calculated proximity of $\bX_{n,m}$ to its closest neighbors, the rate of variation of the phase over neighboring points, or the departure of the amplitude at this point from what would be expected. In cases with small numbers of cores the first of these criteria works very well. However, here we use a much simpler option that consistently gave better results in all our measurements using a larger number of cores. This option simply penalizes points closer to the center of the correlation distribution, where the point density is higher, and favors points near the edge, which tend to be more spaced:
\be
W_{n,m}=\frac{|\bX_{n,m}|}{D},
\ee
where $D$ is the diameter of the core bundle (that is, the maximum value for $|\bX_{n,m}|$), used for normalization purposes. 
The minimization of $\mu$ is achieved by setting to zero its derivatives with respect to each $\phi_n$, leading to the $N$ constraints
\be
\Im\left\{\sum_{m=1}^NP_{n,m}\exp[\ui(\phi_m-\phi_n)]\right\}=0,
\label{Pmult}
\ee
for $n=1,...,N$, where $P_{n,m}=W_{n,m}^2\exp(\ui\Phi_{n,m})$ are the elements of a Hermitian complex matrix $\mathbb{P}$, and $\Im$ denotes the imaginary part. These elements are shown in Fig.~\ref{fig2}(c) for the same realization as in Figs.~\ref{fig2}(a,b). Notice that if $\Phi_{n,m}$ really were equal to the difference of the corresponding phase cores and $W_{n,m}$ were unity, $\mathbb{P}$ would be simply the outer product of a vector with elements $\exp(\ui\phi_m)$ and its own Hermitian conjugate. Therefore, a good initial guess for the unknown phases are the arguments of the leading eigenvector of $\mathbb{P}$. 

A refinement of this initial estimate can be obtained through a rapidly-convergent iterative process. Note that the constraints in Eq.~(\ref{Pmult}) can be written as
\be
\phi_n={\rm arg}\left[\sum_{m=1}^NP_{n,m}\exp(\ui\phi_m)\right]=0,
\label{const}
\ee
where arg denotes the phase. This can be solved iteratively as
\be
\hat{\phi}_n^{(i)}={\rm arg}\left\{\sum_{m=1}^NP_{n,m}\exp[\ui\hat{\phi}_m^{(i-1)}]\right\}.
\ee
That is, a vector whose elements are the phase factors for the $(i-1)^{\rm th}$ iteration is multiplied by $\mathbb{P}$, and the phases of the resulting vector give the phases for the $i^{\rm th}$ iteration. This relation converges rapidly regardless of the choice of $\hat{\phi}_m^{(0)}$, although convergence is greatly accelerated if the initial guess mentioned earlier is used. As can be seen from Fig.~\ref{fig2}(d) for the same realization, this initial estimate is already close to the final result, and after only one iteration the result essentially settles. 
 
\begin{figure}[t] 
\centering 
\includegraphics[scale = .35]{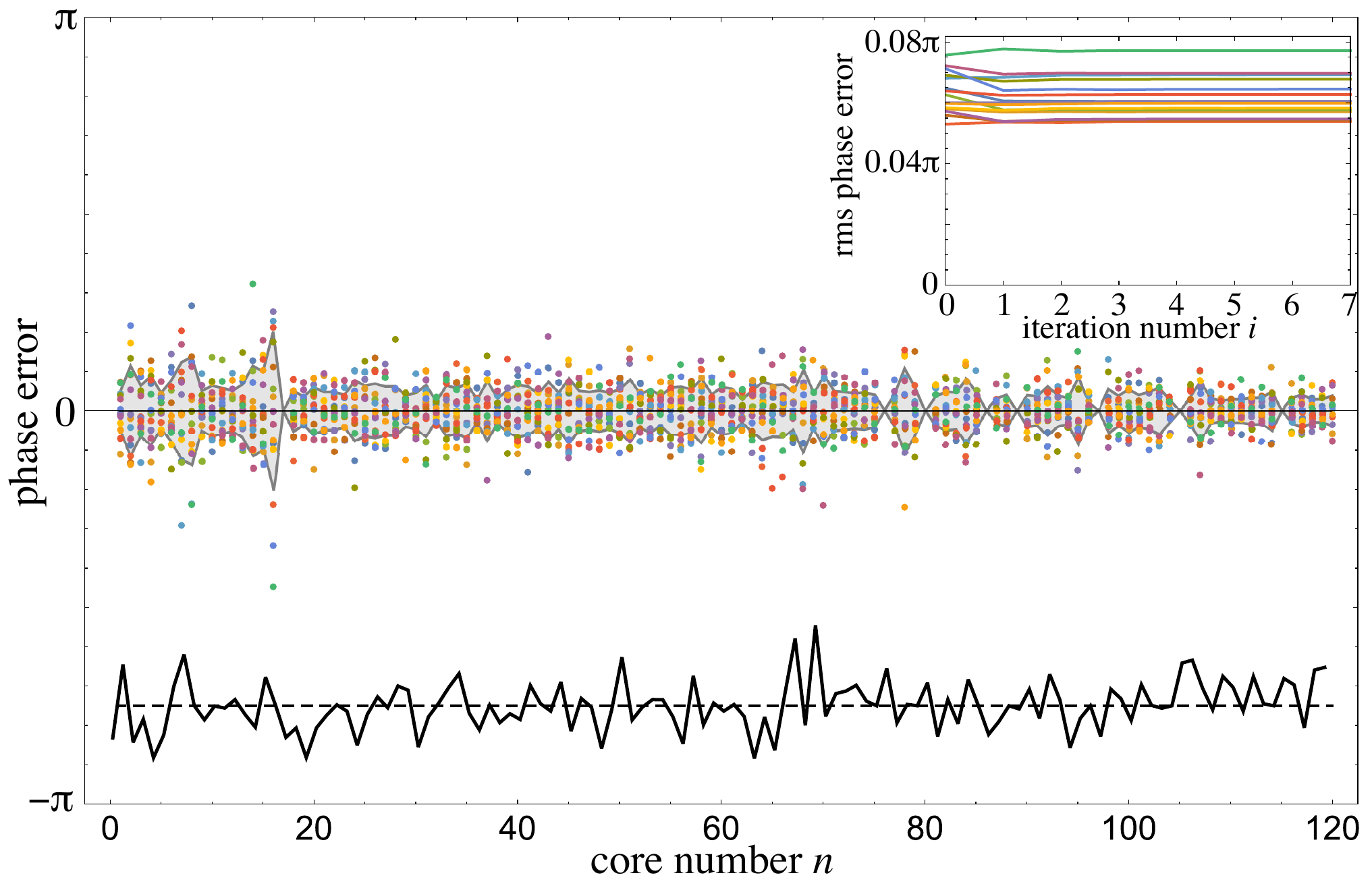} 
\caption{Phase errors from the phase retrieval process. The mismatch between the solid and dashed black lines indicates the systematic phase error for each core. The color dots show the error of the recovered phases once the systematic error is subtracted, where the colors label the 15 realizations. The gray band behind the dots indicates the standard deviation per core. Note that 8 cores are not being used. The inset shows the rms error over all cores (once the systematic error is removed) as a function of iteration number.} 
\label{fig3} 
\end{figure}

The phase retrieval method was tested for 15 independent far-field intensity measurements resulting from applying 15 different sets of  randomly-chosen phases to the cores. These applied phases are known to within the calibration level of error of $~\pi/10$, but for each core, the relative difference between the applied phases is known to within $~\pi/100$, the level of error of the SLM that writes the phase.
It was found that the recovered phases present a systematic error, calculated as the average of the errors for all realizations for each core, and shown as a thick black line (for the used cores) in Fig.~\ref{fig3}. This systematic error with respect to the reference measurements is of the order of $\pi/10$, precisely the level of error of the reference measurements themselves. After subtracting it, the remaining rms error of all the cores for each realization was (with a few exceptions) of a similar scale, as shown by the color dots in Fig.~\ref{fig3}. Overall, the error margin tends to reduce slightly with distance from the core to the center of the MCF, due to the fact that cores near the edge are associated with more non-overlapping correlation points than cores near the center, so that more meaningful constraints apply to them. The inset in Fig.~\ref{fig3} shows for each of the 15 measurements the rms error over all cores as a function of iteration number, which consistently falls below $0.08\pi$. Note that a single iteration brings the error down by about $5\%$ on average, after which the error settles. 

\begin{figure}[b]
  \centering
  \includegraphics[scale=.35]{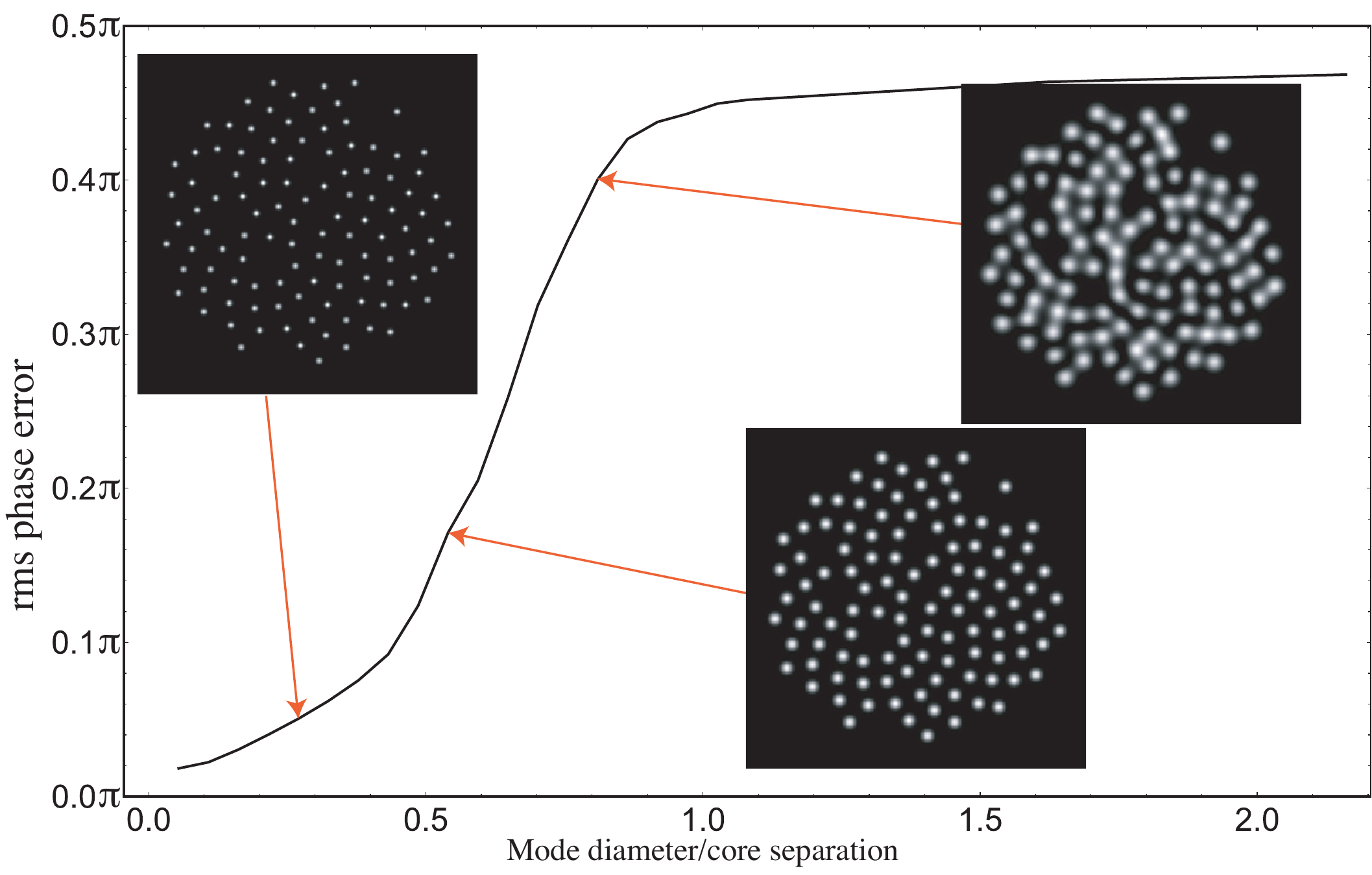}
  \caption{Total rms error of the phase retrieval algorithm over twenty numerical implementations with random phases, as a function of the ratio of mode diameter and core separation. 
\label{fig4}}
\end{figure}

Once the systematic error is removed, the remaining error is not due to the calibration but to the retrieval method, and is directly related to the ratio of mode diameter to core separation. This is confirmed through numerical simulations in which the core diameter was varied. The rms error averaged over twenty randomly chosen sets of phases for 110 cores were calculated as functions of mode diameter. The results are shown in Fig.~\ref{fig4}, where we see that the dependence of this error on the ratio of mode diameter to separation is approximately quadratic until the modes begin to overlap, at which a point the error begins to saturate towards the maximum possible of $\pi/2$. The measured rms error is consistent with the numerical simulations for the corresponding mode diameter/core separation ratio. 

To summarize, we presented a single-shot technique to determine the phases of the transmission matrix of a non-periodic MCF with 112 active cores. The technique makes use of \textit{a priori} information about the core positions and its aperiodicity. 
We observe that the first non-iterative estimate is already close to the actual phases, which is crucial for applications where speed is critical. The iterative process used to refine the result is a $N\times N$ matrix multiplication with low computational cost, which typically converges to a stable solution within a few iterations. The keys to the success of this technique are the aperiodic distribution of cores and their spacing (sufficiently larger than the mode size). 

We underline that all these experiments were performed in transmission. In reflection, the accumulated phase corresponds to a double pass of the MCF system and this brings up the common $\pi$ ambiguity. We note, however, that this ambiguity is not specific to this method, and is also present in interferometric systems \cite{Warren:16a}. Lifting the ambiguity would entail techniques such as spectrally diverse speckle, adding complexity to the system.

While the phase retrieval algorithm and golden spiral arrangement studied here were examined within the context of miniature fiber endoscopes, they can be relevant to other systems employing sparse apertures with small fill factors, such as synthetic aperture imaging and aperture masking interferometry \cite{AMI}, coherent combination of fiber amplifiers in a tiled geometry, or the measurement of coherence \cite{Aura}. If more accurate results are needed, this method can be used to provide good initial estimates for more sophisticated and computationally intensive phase retrieval techniques. 

{\bf Funding.}  
Agence Nationale de la Recherche (ANR-11-EQPX-0017, ANR-10-INSB-04-01, ANR 11-INSB-0006, ANR-14-CE17-0004-01), 
Aix-Marseille Universit\'e (ANR-11-IDEX-0001-02), Universit\'e Lille 1 (ANR-11-LABX-0007), 
Institut National de la Sant\'eet de la Recherche M\'edicale (PC201508), 
SATT Sud-Est (GDC Lensless endoscope), 
CNRS/Weizmann Imaging-Nano European Associated Laboratory. 
G.B. and E.R.A. thank the Ministry of Higher Education and Research, Hauts de France council and European Regional Development Fund (ERDF) through the Contrat de Projets Etat-Region (CPER Photonics for Society P4S). 
M.A.A. acknowledges support from the National Science Foundation (PHY-1507278) and the Excellence Initiative of Aix-Marseille University- A*MIDEX, a French ``Investissements d'Avenir'' program.\\
{\bf Acknowledgements.} We thank Jim Fienup for useful comments.


\begin{thebibliography}{10}
\newcommand{\enquote}[1]{``#1''}

\bibitem{gmitro1993confocal}
A.~F. Gmitro and D.~Aziz, {\protect\JournalTitle{Optics letters}} \textbf{18},
  565 (1993).

\bibitem{gora2013tethered}
M.~J. Gora, J.~S. Sauk, R.~W. Carruth, K.~A. Gallagher, M.~J. Suter, N.~S.
  Nishioka, L.~E. Kava, M.~Rosenberg, B.~E. Bouma, and G.~J. Tearney,
  {\protect\JournalTitle{Nature medicine}} \textbf{19}, 238 (2013).

\bibitem{tearney1996scanning}
G.~Tearney, M.~Brezinski, J.~Fujimoto, N.~Weissman, S.~Boppart, B.~Bouma, and
  J.~Southern, {\protect\JournalTitle{Optics Letters}} \textbf{21}, 543 (1996).

\bibitem{vcivzmar2012exploiting}
T.~{\v{C}}i{\v{z}}m{\'a}r and K.~Dholakia, {\protect\JournalTitle{Nature
  communications}} \textbf{3}, 1027 (2012).

\bibitem{andresen2013toward}
E.~R. Andresen, G.~Bouwmans, S.~Monneret, and H.~Rigneault,
  {\protect\JournalTitle{Optics letters}} \textbf{38}, 609 (2013).

\bibitem{andresen2013two}
E.~R. Andresen, G.~Bouwmans, S.~Monneret, and H.~Rigneault,
  {\protect\JournalTitle{Optics express}} \textbf{21}, 20713 (2013).

\bibitem{Sivankutty2016a}
S.~Sivankutty, E.~R. Andresen, R.~Cossart, G.~Bouwmans, S.~Monneret, and
  H.~Rigneault, {\protect\JournalTitle{Opt. Express}} \textbf{24}, 825 (2016).

\bibitem{psaltis_multimode}
I.~N. Papadopoulos, S.~Farahi, C.~Moser, and D.~Psaltis,
  {\protect\JournalTitle{Biomed. Opt. Express}} \textbf{4}, 260 (2013).

\bibitem{choi_multimode}
Y.~Choi, C.~Yoon, M.~Kim, T.~D. Yang, C.~Fang-Yen, R.~R. Dasari, K.~J. Lee, and
  W.~Choi, {\protect\JournalTitle{Phys. Rev. Lett.}} \textbf{109}, 203901
  (2012).

\bibitem{ploschner2015multimode}
M.~Pl{\"o}schner, V.~Koll{\'a}rov{\'a}, Z.~Dost{\'a}l, J.~Nylk, T.~Barton-Owen,
  D.~E. Ferrier, R.~Chmel{\'\i}k, K.~Dholakia, and T.~{\v{C}}i{\v{z}}m{\'a}r,
  {\protect\JournalTitle{Scientific reports}} \textbf{5} (2015).

\bibitem{thompson2011adaptive}
A.~J. Thompson, C.~Paterson, M.~A. Neil, C.~Dunsby, and P.~M. French,
  {\protect\JournalTitle{Optics letters}} \textbf{36}, 1707 (2011).

\bibitem{Sivankutty:16}
S.~Sivankutty, V.~Tsvirkun, G.~Bouwmans, D.~Kogan, D.~Oron, E.~R. Andresen, and
  H.~Rigneault, {\protect\JournalTitle{Opt. Lett.}} \textbf{41}, 3531 (2016).

\bibitem{Tsvirkun:17}
V.~Tsvirkun, S.~Sivankutty, G.~Bouwmans, O.~Vanvincq, E.~R. Andresen, and
  H.~Rigneault, {\protect\JournalTitle{Opt. Express}} \textbf{25}, 31863
  (2017).

\bibitem{Sivankutty:18a}
S.~Sivankutty, V.~Tsvirkun, O.~Vanvincq, G.~Bouwmans, E.~R. Andresen, and
  H.~Rigneault, {\protect\JournalTitle{Opt. Lett.}} \textbf{43}, 3638 (2018).

\bibitem{gabrielli2016aperiodic}
L.~H. Gabrielli and H.~E. Hernandez-Figueroa, {\protect\JournalTitle{IEEE
  Photonics Technology Letters}} \textbf{28}, 209.

\bibitem{caravaca2013real}
A.~M. Caravaca-Aguirre, E.~Niv, D.~B. Conkey, and R.~Piestun,
  {\protect\JournalTitle{Optics Express}} \textbf{21}, 12881 (2013).

\bibitem{Kogan:17}
D.~Kogan, S.~Sivankutty, V.~Tsvirkun, G.~Bouwmans, E.~R. Andresen,
  H.~Rigneault, and D.~Oron, {\protect\JournalTitle{Opt. Lett.}} \textbf{42},
  647 (2017).

\bibitem{fienup_review}
J.~R. Fienup, {\protect\JournalTitle{Appl. Opt.}} \textbf{21}, 2758 (1982).

\bibitem{dremeau2015reference}
A.~Dr{\'e}meau, A.~Liutkus, D.~Martina, O.~Katz, C.~Sch{\"u}lke, F.~Krzakala,
  S.~Gigan, and L.~Daudet, {\protect\JournalTitle{Optics express}} \textbf{23},
  11898 (2015).

\bibitem{AMI}
P.~G. Tuthill, J.~D. Monnier, and W.~C. Danchi, vol. 4006, pp.
  491--499.

\bibitem{Sivankutty:16b}
S.~Sivankutty, E.~R. Andresen, G.~Bouwmans, T.~G. Brown, M.~A. Alonso, and
  H.~Rigneault, {\protect\JournalTitle{Opt. Lett.}} \textbf{41}, 2105 (2016).

\bibitem{Warren:16a}
S.~C. Warren, Y.~Kim, J.~M. Stone, C.~Mitchell, J.~C. Knight, M.~A.~A. Neil,
  C.~Paterson, P.~M.~W. French, and C.~Dunsby, {\protect\JournalTitle{Opt.
  Express}} \textbf{24}, 21474 (2016).

\bibitem{Aura}
Y.~Mej{\'\i}a and A.~I. Gonz{\'a}lez, {\protect\JournalTitle{Optics
  Communications}} \textbf{273}, 428 (2007).

\end{thebibliography}
\end{document}